\documentclass[conference]{IEEEtran}
\IEEEoverridecommandlockouts
\usepackage{cite}
\usepackage{amsmath,amssymb,amsfonts}
\usepackage{algorithmic}
\usepackage{graphicx}
\usepackage{textcomp}
\usepackage{xcolor}
\usepackage{wrapfig}
\usepackage{textcomp}
\usepackage{xcolor}
\usepackage{soul}
\usepackage{braket}
\usepackage{algorithmic}
\usepackage[ruled,vlined, linesnumbered]{algorithm2e}
\usepackage{graphicx}
\usepackage{cite}
\usepackage{amsmath,amssymb,amsfonts}
\usepackage{algorithmic}
\usepackage{graphicx}
\usepackage{textcomp}
\usepackage{xcolor}
\usepackage{wrapfig}
\usepackage{textcomp}
\usepackage{xcolor}
\usepackage{soul}
\usepackage{braket}
\usepackage{algorithmic}
\usepackage[ruled,vlined, linesnumbered]{algorithm2e}
\usepackage{graphicx}
\usepackage{cite}
\usepackage{amsmath,amssymb,amsfonts}
\usepackage{algorithmic}
\usepackage{graphicx}
\usepackage{textcomp}
\usepackage{xcolor}
\usepackage{multicol}
\usepackage{subcaption}

\def\BibTeX{{\rm B\kern-.05em{\sc i\kern-.025em b}\kern-.08em
    T\kern-.1667em\lower.7ex\hbox{E}\kern-.125emX}}
\begin{document}
{
\title{Designing Hash and Encryption Engines using Quantum Computing\\
 }

 \author{\IEEEauthorblockN{Suryansh Upadhyay}
\IEEEauthorblockA{\textit{The Pennsylvania State University} \\
\textit{University Park, PA, USA} \\
sju5079@psu.edu}
\and
\IEEEauthorblockN{Rupshali Roy}
\IEEEauthorblockA{\textit{The Pennsylvania State University} \\
\textit{University Park, PA, USA} \\
rzr5509@psu.edu}

\and
\IEEEauthorblockN{Swaroop Ghosh}
\IEEEauthorblockA{\textit{The Pennsylvania State University} \\
\textit{University Park, PA, USA} \\
szg212@psu.edu}
}


\maketitle

\begin{abstract}

Quantum computing (QC) holds the promise of revolutionizing problem-solving by exploiting quantum phenomena like superposition and entanglement. It offers exponential speed-ups across various domains, from machine learning and security to drug discovery and optimization. In parallel, quantum encryption and key distribution have garnered substantial interest, leveraging quantum engines to enhance cryptographic techniques. Classical cryptography faces imminent threats from quantum computing, exemplified by Shor's algorithm's capacity to breach established encryption schemes. However, quantum circuits and algorithms, capitalizing on superposition and entanglement, offer innovative avenues for enhancing security. In this paper we explore quantum-based hash functions and encryption to fortify data security. Quantum hash functions and encryption can have numerous potential application cases, such as password storage, digital signatures, cryptography, anti-tampering etc. 
The integration of quantum and classical methods demonstrates potential in securing data in the era of quantum computing.
\end{abstract}


\section{Introduction}

Quantum computing (QC) has become the focus of extensive research and attention, driven by its potential to revolutionize problem-solving across diverse domains. When tackling combinatorial problems, quantum computers can achieve exponential speedups by harnessing quantum-mechanical phenomena such as superposition and entanglement. The broad spectrum of applications for quantum computing spans areas such as machine learning\cite{b1}, security\cite{b2}, drug discovery\cite{b3} etc. However, the practical realization of quantum computing face significant obstacles, such as qubit decoherence, measurement inaccuracies, gate errors, and temporal fluctuations. Quantum error correction (QEC) codes\cite{b5} present a promising solution for ensuring reliable quantum operations. Still, their current demand for substantial resources makes them impractical for widespread adoption in the immediate future. The emergence of Noisy Intermediate-Scale Quantum (NISQ) computers, which have a limited number of qubits and operate in the presence of noise, along with various hybrid algorithms offer a potential solution to important problems such as discrete optimization, quantum chemical simulations and drug discovery. 

In the wake of recent advancements in quantum computing, there has been a notable surge of interest in the domains of quantum encryption and quantum key distribution. These developments have ignited intriguing possibilities such as , use of quantum algorithms for developing robust and efficient cryptographic primitives \cite{b6}\cite{b7} that form the backbone of secure data communication and storage.

\textbf{Motivation:} The advent and rapid progress in quantum computing have ushered in a new era that both threatens and inspires innovation in the realm of cryptographic codes and data security. \textbf{a) Threat to classical cryptography:} Quantum computing's computational power poses threat to classical cryptographic codes. Shor's algorithm, for example, has the potential to break widely-used encryption schemes like RSA by efficiently factoring large numbers \cite{b11}. This capability undermines the security foundations of much of today's digital communication and data protection. \textbf{b) The need for quantum-resistant solutions:} As quantum computers advance, many classical encryption methods will become obsolete hence there is an urgent need for cryptographic systems that are resilient to quantum attacks. Developing and deploying quantum-resistant cryptographic codes is a technical imperative in safeguarding sensitive information in a quantum-powered world. \textbf{c) Quantum algorithms as a solution:} However, in the face of this quantum challenge, there lies an opportunity for innovation and enhanced security. Quantum circuits and algorithms, with their unique properties like superposition and entanglement, can be used to design cryptographic solutions that might be more secure against attacks while also being more robust in terms of performance.

In this work we explore the potential of quantum-based hash function and encryption algorithm in fortifying data security. Quantum hash functions can leverage quantum properties to enhance data integrity verification and collision resistance, addressing vulnerabilities exposed by quantum computing. It can serve as a robust tool for verifying the integrity of data and safeguarding against potential adversarial threats. For instance, a user can utilize a quantum hash function to create a distinctive hash value for their data. Once the data is secured using this quantum hash function, it is transmitted to the recipient. Alongside the data, a seed that specifies the specific hash function used is also shared. The recipient can then verify the incoming data by applying the same hash function. Furthermore, the quantum hash function's inherently chaotic dynamics may enable it's use in various other applications, including the generation of pseudo-random numbers and the development of image encryption algorithms based on quantum hash functions.

State-of-the-art implementations of quantum AES offer improved quantum cryptanalysis estimations, i.e. they show resistance against quantum algorithms such as Shor's algorithm or Grover's algorithm, unlike existing classical encryption standards. However, our objective is to show that encryption can be done in a completely different way (and by following completely different steps/algorithms) using quantum computing. It can potentially be scaled to larger dimensions and evaluated for resistance to quantum cryptalysis however, the present study is limited to conceptual level. As such, we have not compared the proposed design with existing large scale quantum AES or delved into the cryptanalysis estimations.

The quantum-enhanced encryption protocol utilizes a lookup table and a matrix operator (which is applied on qubits in the form of a quantum circuit) for data encryption. The sender encrypts the binary data, resulting in ciphertext (represented as measured qubits to the receiver). To successfully decrypt the data, the receiver also receives a seed that contains the lookup table and matrix operator used in encryption. Since the lookup table is the inverse of itself, the receiver applies the inverse of the matrix and lookup table in reverse order. It is important to ensure that equally efficient hardware/backends, having the same coupling map are used by both the sender and the receiver for the encryption/decryption process, so that the matrix operator and its inverse translate into the desired quantum circuit and its reverse without extra number of gates and circuit depth (which in turn will maintain similar level of noise in encryption and decryption process). 

\textbf{Paper organization:} Section II provides background information on quantum computing, terms used and related work. Section III discusses the quantum based hash functions. In section IV we present a quantum based implementation of AES-128. Section V concludes the paper.

\section{Background}


\subsection{Qubits and Quantum gates}

Qubits due to their quantum nature, can exist in a superposition of both $\ket{0}$ and $\ket{1}$ unlike classical bits. As a result, a n-qubit system can represent all $2^n$ basis states at the same time. A qubit's state, denoted by $\varphi$ = a $\ket{0}$ + b $\ket{1}$, can be expressed as a combination of complex probability amplitudes, $a$ and $b$, corresponding to the states $\ket{0}$ and $\ket{1}$. 
When a qubit is measured, it collapses to a single state, either $\ket{0}$ or $\ket{1}$, with probabilities of $|a|^2$ and $|b|^2$, respectively. Furthermore, qubits can be entangled, which means that the states of multiple qubits become correlated. 
To perform calculations effectively, quantum computation relies on the manipulation of qubit states. A quantum program performs a series of gate operations on a group of correctly initialized qubits (using laser pulses in ion trap qubits and RF pulses in superconducting qubits). 
Mathematically, quantum gates are represented using unitary matrices (a matrix U is unitary if U$U^\dagger$ = I, where U$^\dagger$ is the adjoint of matrix U and I is the identity matrix)



\subsection{Parameterized quantum circuits (PQC)}

PQC is built from a collection of parameterized and controlled single qubit gates. A classical optimizer optimizes the parameters iteratively to achieve the desired input-output relationship. A PQC is used by a quantum processor to prepare a quantum state. The classical computer generates a new set of optimized parameters for the PQC based on the output distribution, which is then fed back to the quantum computer. The entire procedure continues in a closed loop until a traditional optimization target is reached. 

\subsection{Hash function}

A hash function is a mathematical function that converts variable-length data into fixed-length values, though some hash functions can also produce variable-length outputs. A hash function's output is commonly referred to as hash values, hash codes, digests, or simply hashes. These hash values are typically used to index a hash table, which is a fixed-size data structure. A hash function accepts a key as input to uniquely identify a datum or record within a data storage. These keys can be fixed-length, such as integers, or variable-length, such as names. In some cases, the key itself may represent the datum. 
This process yields a hash code, which is used to effectively index a hash table containing the data or records, or references to them. 

\begin{figure}
    \centering
    \includegraphics[width= 3.5in]{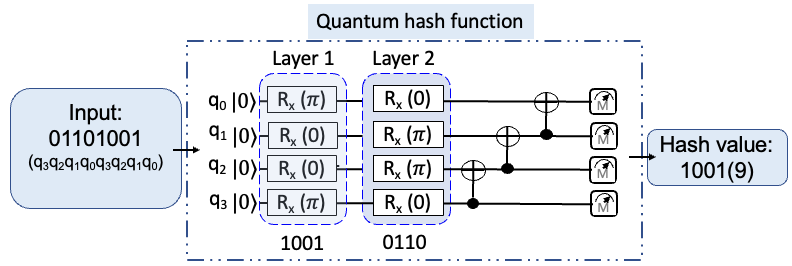}
    \caption{Generating a 4-bit numerical hash value from an 8-bit input bit string. Encoding occurs in two layers within the circuit: the first layer encodes the first four bits using rotation angles, employing $\pi$ radians for bit value 1 and 0 radians for bit value 0. The subsequent layer encodes the remaining four bits.}
    \label{1}
\vspace{-4mm}
\end{figure}


\begin{figure*}
    \centering
    \includegraphics[width= 7.15in]{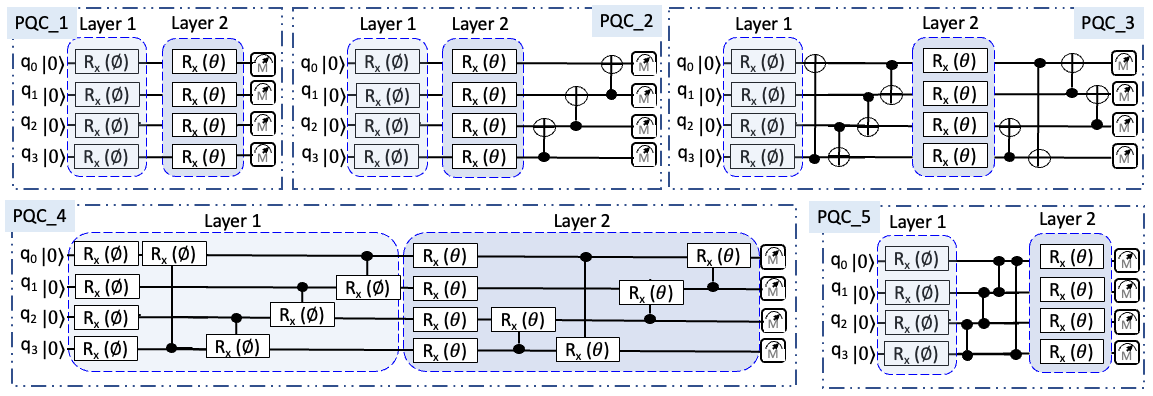}
    \caption{Different PQC circuits explored as candidates for hash functions.}
    \label{2}
\vspace{-4mm}
\end{figure*}

\subsection{Encryption}

Advanced Encryption Standard (AES), is a popular symmetric encryption algorithm for data and communications security. This encryption method uses a (AES-n) n-bit key, which means it encrypts and decrypts data using n bits of data as the secret key. AES-128 is widely used in a variety of applications, including secure communication protocols, data encryption, and data protection mechanisms in modern computing systems, due to its efficiency and robust security features.

\subsection{Related work}

Several studies have focused on creating optimal quantum circuits for block ciphers. Kim et al. proposed optimal quantum circuits for SHA-2 based on its message expansion function in \cite{b16}, while Song et al. presented a new quantum circuit implementation for SM3 in \cite{b17}. These studies demonstrate that optimal classical circuit designs can be adapted into optimal quantum circuits, although they face challenges such as high depth and error rates. In our work, we investigate well-established and error-tolerant PQCs as potential hash functions. Quantum implementations of AES have emerged, notably by Grassl et al. \cite{b7}, refined by Kim et al. \cite{b8}, and improved by Langenberg et al. \cite{b9}, reducing quantum gate requirements. The AES has proven its resistance to quantum attacks with the costs of doubling key sizes \cite{b6}. Wang et al. presented an efficient quantum AES-128 implementation \cite{b10}, demanding fewer gates and qubits. Kuang and Bettenburg introduced the Quantum Permutation Pad (QPP) offering a versatile symmetric encryption solution for both quantum and classical systems \cite{b12}. While QPP's unconventional gate permutations pose encryption strength questions, this paper presents a proof-of-concept for implementing AES-128 steps (SubBytes, MixColumns and ShiftRows) on 4-qubit data chunks, showcasing encryption and decryption transformations on a reduced-size image. 

\begin{figure}
    \centering
    \includegraphics[width= 3.5in]{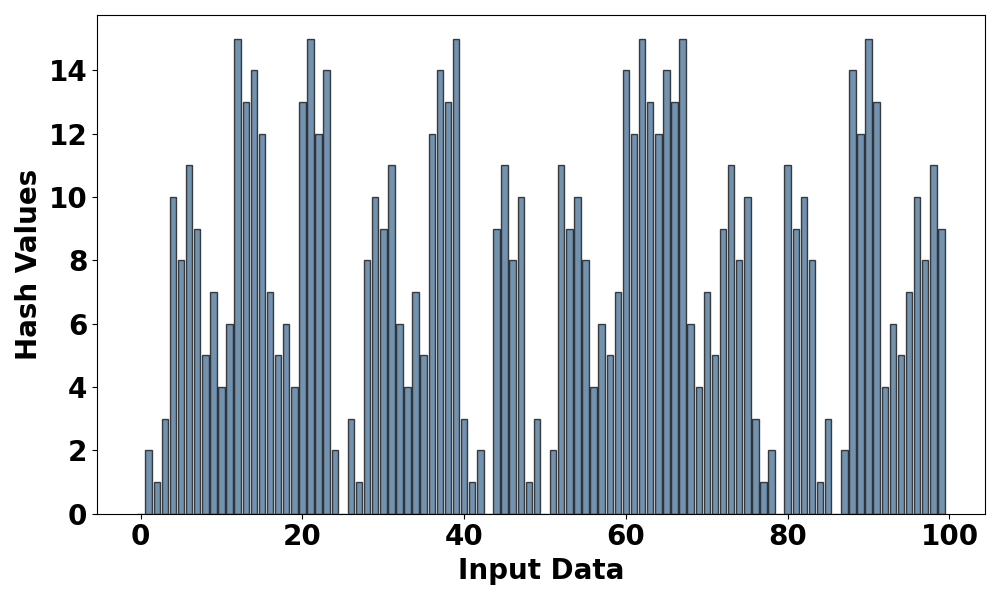}
    \caption{Validation of PQC-3 as a quantum hash function on fake backend (fake\_vigo[6 qubit]). A 4-qubit PQC circuit variant processes a limited 100-input batch of 8-bit bitstrings (0 to 99) with 1000 shots each.}
    \label{3}
\vspace{-4mm}
\end{figure}

\section{Quantum hash function}


\subsection{Basic Idea}

In the context of quantum hash functions, it is important to note that there can be an extensive number of approaches and circuits that can potentially serve as effective hash functions. In this work, we focus on PQCs. These PQCs are employed to encode input data using rotation angles, ultimately generating a hash string or hash value as the output. An essential consideration in this implementation is the quantum circuit's ability to effectively and uniformly address the Hilbert space as this influences the distribution and quality of hash values generated by the quantum circuit and, consequently, the overall performance of the quantum hash function. PQCs offer the advantage of high-dimensional Hilbert spaces, potentially improving accuracy in hash function generation. To find a dependable quantum hash function, we adhere to a set of critical properties that a hash function must have. These characteristics serve as benchmarks for assessing the quality and efficacy of any hash function:

a) \textbf{Deterministic:} it consistently produces the same hash value for a given input, ensuring predictability.

b) \textbf{Fixed Size Output:} hash values are of a set size, simplifying handling.

c) \textbf{Efficient:} swift hash value generation supports real-time applications.

d) \textbf{Pre-image Resistance:} prevents reverse-engineering of the input from its hash value, enhancing security.

e) \textbf{Collision Resistance:} makes finding two inputs with the same hash value highly challenging, ensuring data integrity.

f) \textbf{Avalanche Effect:} small input changes yield significantly different hash values, bolstering security.

g) \textbf{Uniform Distribution:} hash values are evenly distributed across the entire range for clustering prevention.

\begin{algorithm}[t]
\caption{Quantum hash function}
\label{algo:bitstring_to_weights}
\SetAlgoLined
\KwIn{Input (image/integer/bit string); PQC circuit}
\KwOut{Quantum hash function with encoded inputs}
\begin{enumerate}
    \item Convert input to binary bit string.
    \item Loop through each bit in the binary bit string (\textit{bit\_string}).
    \begin{enumerate}
        \item For the first $\frac{n}{2}$ bits:
        \begin{enumerate}
            \item If bit $\equiv$ '1', set Rx gate on qubit n (\textit{Layer 1}) to $\theta_1$.
            \item If bit $\equiv$ '0', set Rx gate on qubit n  (\textit{Layer 1}) to 0 (or $\phi_1$ if needed).
        \end{enumerate}
        \item For the last $\frac{n}{2}$ bits:
        \begin{enumerate}
            \item If bit $\equiv$ '1', set Rx gate on qubit n (\textit{Layer 2}) to $\theta_2$.
            \item If bit $\equiv$ '0', set Rx gate on qubit n (\textit{Layer 2}) to 0 (or $\phi_2$ if needed).
        \end{enumerate}
    \end{enumerate}
    \item \textbf{Output:} Hash values of the input.
\end{enumerate}
\end{algorithm}

\subsection{Implementation}

To ensure the robustness and reliability of our approach, we leverage well-established PQCs that have been extensively studied in the field of quantum computing\cite{b15}. Our approach involves the encoding of input data into these PQCs by representing the input as rotation angles for the circuit's rotation gates. We present a generalized encoding approach in Algorithm 1. Any data can be transformed into bit strings, the bit value within the bit string determines whether it is encoded using an angle denoted as $\theta_1$ (for a bit value of 1) or $\phi_1$ (for a bit value of 0) within the rotation gate. For example: let's consider an input bitstring: 1001. In this case, the rotation values for the rotation gates applied to the qubits $(q_3, q_2, q_1, q_0)$ will be as follows: $\theta_1$, $\phi_1$, $\phi_1$, $\theta_1$.

Fig. \ref{1} illustrates the process of generating a 4-bit numerical hash value from an 8-bit input bit string. In this particular scenario, we begin by setting all qubits to the 0 state. Within the circuit, the encoding process takes place in two distinct layers. We encode the first four bits of the input bit string as rotation angles in the first layer. If the bit has a value of 1, we use $\pi$ radians as the corresponding angle, and if the bit has a value of 0, we use 0 radians. The remaining four bits of the input bit string are then encoded in the following layer of the circuit. The values obtained from measurements form the resulting hash value.

\subsection{Result and analysis}

\subsubsection{Experimental setup}

We leverage the Qiskit open-source quantum software development kit from IBM, employing a Python wrapper for simulations. For \textbf{benchmarks}, we make use of different PQCs\cite{b15} Fig.\ref{2}. For \textbf{benchmark execution}, we utilize Qiskit's fake provider module (fake\_vigo[6 qubit], fake\_singapore[20 qubit]), which comprises noisy simulators mimicking real IBM Quantum systems through system snapshots. These snapshots contain crucial information about the quantum system, such as the coupling map, basis gates, and qubit parameters. 
For \textbf{performance metric} we use:

\begin{figure}
    \centering
    \includegraphics[width= 3.5in]{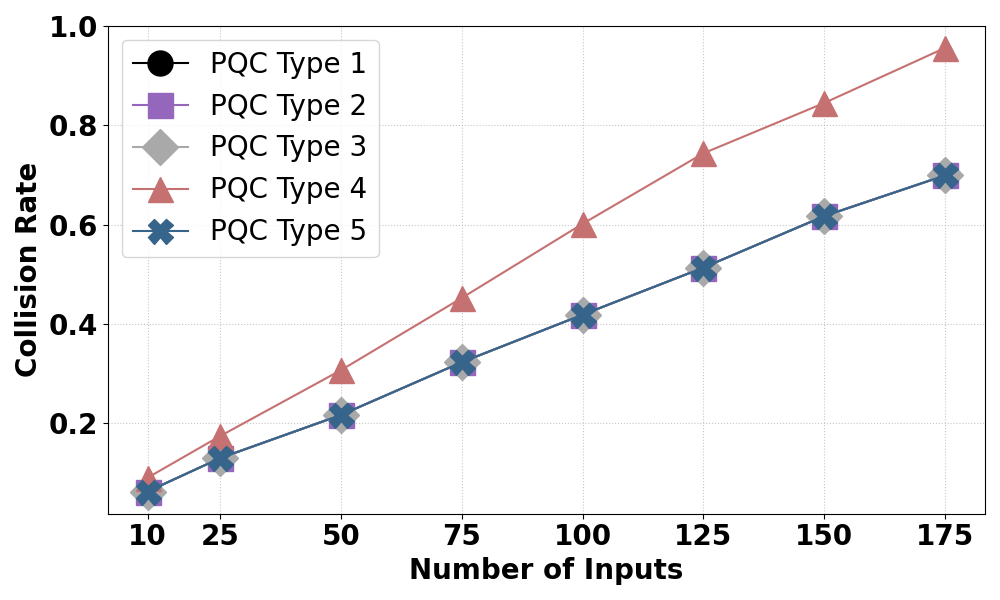}
    \caption{The 4-qubit variants of various PQC circuits and their collision response to different batch sizes. Run on fake\_vigo for 1000 shots each.}
    \label{4}
\vspace{-4mm}
\end{figure}

\begin{figure}
    \centering
    \includegraphics[width= 3.5in]{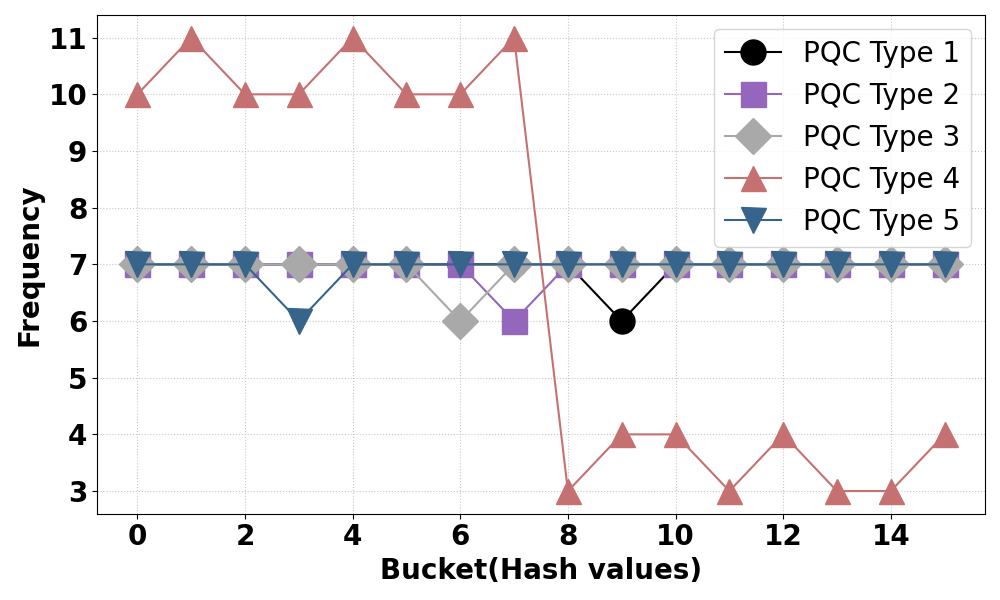}
    \caption{Bucket histogram for the proposed PQC hash functions. A bucket represents an output hash value. We use 4-qubit PQCs to process a batch of 100 inputs represented as 8-bit bitstrings, run on fake\_vigo for 1000 shots each.}
    \label{5}
\vspace{-4mm}
\end{figure}

1) \textbf{Collision rate (CR):} serves as a metric for quantifying collisions in a quantum hash function when applied to a particular input dataset. It is defined as-

\begin{equation}
\text{CR} = \frac{\text{$f_{avg} + stdev$}}{2^{\text{qubits}}}
\end{equation}

Here, $f_{avg}$ and $stdev$ denotes the average frequency of hash values and standard deviation generated by a specific function for a given input data-set, and $2^{qubits}$ represents the total count of distinct possible hash values attainable by an n-qubit quantum hash circuit. A \textit{lower collision rate indicates superior performance} of the quantum hash function when applied to the given input values.

2) \textbf{Buckets histograms:} provide insights into the distribution of hash values among various buckets and reveal the frequency of hash value recurrence within a given input set. These histograms serve as a valuable tool for identifying anomalous patterns. \textit{In an ideal scenario, hash values should exhibit uniform distribution, indicating that they appear with equal frequency across the entire range} to mitigate clustering.

3) \textbf{Statistical goodness test:} The chi-squared test is used to ascertain whether an observed probability distribution aligns with a known and expected distribution. In our specific context, our objective is to assess whether the distribution of output hash values across all possible outputs follows a uniform and random pattern. The p-value derived from this test quantifies the likelihood that the behavior of the hash function resembles that of a uniformly distributed random variable or not. \textit{A p-value close to 1 indicates strong hash function performance, signifying that the observed distribution is not statistically significantly different from the expected distribution.}

\subsubsection{Concept validation}

We validate PQC's ($PQC\_3$) application as a hash function using a fake backend (fake\_vigo) Fig.\ref{3}. A 4-qubit variant of the PQC circuit is used, processing 100 input batch (due to limited hardware availability) of 8-bit bitstrings (ranging from 0 to 99) run for 1000 shots each. For simplicity the values of $\theta$, $\phi$ are either 0 or $\pi$, depending on the input bit being encoded ( $\pi$ if bit is 1, else 0). The figure demonstrates the generation of well-distributed hash values. This empirical validation highlights the practicality of the PQC circuit as a hash function.

\subsubsection{Performance evaluation}

We assess the performance of the proposed PQC hash functions using a bucket histogram, as depicted in Fig. \ref{5}. We use 4-qubit PQCs, processing a batch of 100 inputs (8-bit bitstrings). Notably, PQCs 1, 2, 3, and 5 exhibit robust performance, as their hash values are evenly distributed across the entire range. In contrast, PQC$\_4$ displays a non-uniform frequency distribution. Furthermore, we have computed the p-values for each circuit: \textit{PQC$\_1$: 1, PQC$\_2$: 1, PQC$\_3$: 1, PQC$\_4$: 0.02, PQC$\_5$: 1}. A p-value close to 1 indicates a strong hash function performance, suggesting that the observed distribution closely aligns with the expected distribution, with the exception of PQC$_4$, which shows statistically significant deviation.

\subsubsection{Batch size impact on collisions}

The effect of batch size on collision rates is a critical consideration in evaluating the performance of hash function. We compare the 4 qubit variant of each PQC circuit and how they handle different batch sizes Fig. \ref{4}. For simplicity $\theta$ and $\phi$ take on either 0 or $\pi$. As the batch size increases, the collision rate tends to rise proportionally. When batch sizes are expanded, more data inputs are processed simultaneously. This increased input volume introduces a higher probability of two or more inputs coincidentally generating identical hash values, thus elevating the collision rate. We notice consistent collision performance across PQC 1, 2, 3, and 5, with PQC$\_$4 exhibiting the highest collision rate among them.

\subsection{Discussion}


In PQC, the choice of encoding angles is crucial as it affects how points are distributed in hilbert space. In this work, we use angles of 0 and $\pi$ to represent input bits 0 and 1, but there can be various ways to encode data. An interesting extension can be assigning different angles to input bits based on a weighting scheme, offering more precise control over how the output hash values are distributed. For instance, we could weigh the angles based on the probability distribution of input bit values, potentially improving the hash function's performance (collision resistance, cluster prevention), especially for data with non-uniform bit distributions. Noise can substantially affect the reliability and precision of quantum operations. Moreover, it can perturb the deterministic nature of quantum hash functions. Judicious selection of quantum circuit and encoding methodologies are instrumental in mitigating the effects of noise.

\section{Quantum AES implementation}

\subsection{Methodology and Results}
We use the Qiskit for simulations and fake provider module (fake\_valencia[5 qubit]) for \textbf{benchmark execution}.
Our implementation adapts the \textit{SubBytes}, \textit{MixColumns}, \textit{ShiftRows} steps of the classical AES-128 to a mix of classical and quantum operations to encrypt and decrypt data (Fig. \ref{fig:illustration}). We illustrate the process using a 10X10 image of the alphabet A as data input (Fig. \ref{fig:illustration}). 

\begin{figure*}
    \centering
    \includegraphics[width= 7.15in]{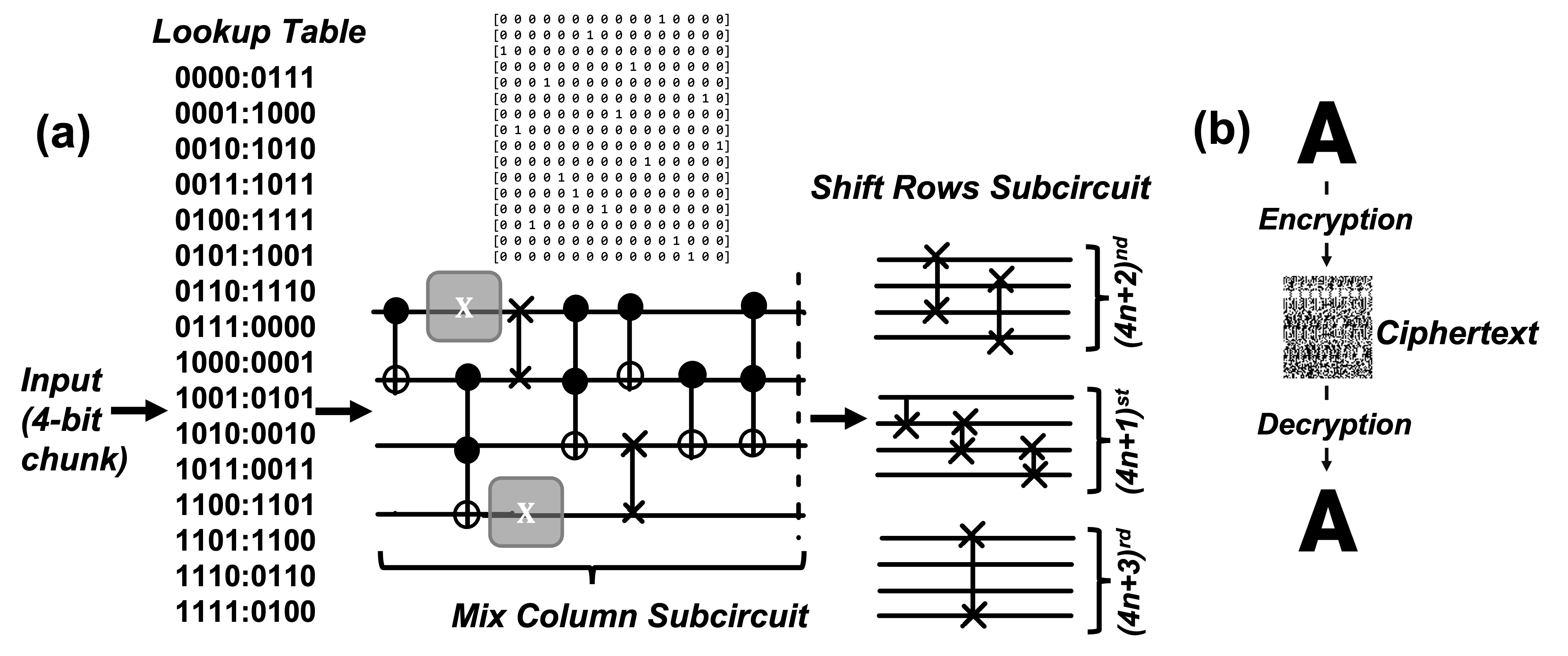}
    \caption{a) Schematic of the circuitry used to encrypt the binary chunks from the image. b) Input image of the alphabet 'A', (reduced to $10 \times 10$ pixel), encrypted into a cipher image, then decrypted back into the original image.}
    \label{fig:illustration}
\vspace{-4mm}
\end{figure*}

\textbf{Encryption:} We first convert the image into a binary bit string and segment it into 4-bit units. For \textit{SubBytes} substitution, a classical 16-entry lookup table is employed, replacing each 4-bit segment with the corresponding entry. Next, these 4-bit units are processed through \textit{MixColumns}. To apply a matrix to a circuit with $n$ (4 in our case) qubits, we need a matrix of dimensions $2^n$ $\times$ $2^n$, i.e. $16$ $\times$ $16$. Since this is a preliminary attempt at transforming the \textit{MixColumns} step in the quantum domain, we have used a real matrix $[D]$ to the qubits. When used as an operator in the circuit, the matrix translates into a series of classical gates like controlled-NOT, controlled-CNOT, X and SWAP gates. Following this, \textit{ShiftRows} is executed using SWAP gates, where specific bit segments undergo left circular shifts. Every $(4n+1)^{st}$ chunk undergoes a left circular shift by one position, while every $(4n+2)^{nd}$ chunk undergoes the shift by 2 positions, every $(4n+3)^{rd}$ chunk undergoes the shift by 3 positions. Every $(4n)^{th}$ chunk is left unchanged. The resulting processed bit segments are concatenated to generate a binary ciphertext. This ciphertext effectively conceals the original image data, rendering it indecipherable as an image file. To evaluate the encryption's impact, the binary string is examined using a raw pixel viewer.

\textbf{Decryption:} Each decryption step is the inverse of its corresponding encryption step, applied in reverse order. We start by applying the inverse of \textit{ShiftRows} which uses a right circular shift on the bits. To implement this we use the SWAP gates in the reverse order compared to encryption. Subsequently, we execute the inverse of \textit{MixColumns} by applying the matrix operator $[D]^{-1}$ to the qubits. This corresponds to the same sequence of gates utilized in the encryption circuit, but in reverse order. Finally, to undo the \textit{SubBytes} step, each chunk of 4 bits is substituted by its corresponding values from the lookup table we used during the \textit{SubBytes} step in encryption. This lookup table is designed to be self-inverse, enabling its utilization for decryption. The resulting values are concatenated into a binary bitstring, which is then transformed back into an image file, effectively restoring the original input. This successful decryption process demonstrates the reversibility of the encryption (Fig. \ref{fig:illustration}).

\subsection{Drawbacks and Scope for Improvement}

The matrix employed for the \textit{MixColumns} step is expressed in the form of CCX, CX, X, and SWAP gates within a corresponding quantum circuit. This representation implies that the qubits remain free from superposition or phase alterations. As a result, when we measure both the ciphertext qubits and the final decrypted qubits, we consistently observe a single basis state with the highest frequency, thereby achieving a pure state. This characteristic translates to zero entropy, as there are no intrinsic uncertainties involved in the measurements. The allowable matrices could be expanded to include complex matrices, which would correspond to gates involving superposition and/or phase shifts. In such a scenario, measuring the ciphertext qubits would yield multiple basis states with the highest frequency for several 4-bit chunks, resulting in higher entropy. We could replace the lookup table used in the \textit{SubBytes} step with a quantum read-only memory (QROM), to bring this step to the quantum domain from the classical domain. Additionally, our model does not incorporate the \textit{AddRoundKey} step from AES, indicating that we do not use a key. However, this could be a plausible addition to enhance the model in the future.

\section{Conclusion}

Quantum computers pose a significant threat to classical encryption methods, demanding the development of quantum-resistant cryptography. Yet, quantum properties like superposition and entanglement offer a chance to innovate in cryptography. In this work we've explored the potential of quantum-based hash functions and AES algorithms for data security. Data integrity and collision resistance may be enhanced by quantum hash functions. Cryptography with quantum enhancements may guarantee stronger encryption and decryption, protecting sensitive data. 

\section{Acknowledgment}

This work is supported in parts by NSF (CNS-1722557, CNS-2129675, CCF-2210963, CCF-1718474, OIA-2040667, DGE-1723687, DGE-1821766, and DGE-2113839), Intel's gift and seed grants from Penn State ICDS and Huck Institute of the Life Sciences.


\end{document}